\documentclass[11pt,reprint,english,aps,prc,showpacs,letter,longbibliography]{revtex4-1}

\usepackage[     bookmarks,
                 bookmarksopen = true,
                 bookmarksnumbered = true,
                 linktocpage,
                 colorlinks = true,
                 linkcolor = blue,
                 urlcolor  = blue,
                 citecolor = blue,
                 anchorcolor = green,
                 hyperindex = true,
                 hyperfigures]
        {hyperref}

\usepackage{graphicx}
\usepackage{dcolumn}
\usepackage{bm}
\usepackage{float}
\usepackage{amsmath}

\usepackage{color}
\usepackage{graphicx}
\definecolor{dgreen}{cmyk}{1.,0.,1.,0.2}        
\definecolor{orange}{cmyk}{0.,0.353,1.,0.}    

\begin{document}

\title{Deep-learning quasi-particle masses from QCD equation of state}%

\author{Fu-Peng Li$^{1}$, Hong-Liang L\"u$^2$, Long-Gang Pang$^{1*}$\footnote{email: lgpang@ccnu.edu.cn}, 
Guang-You Qin$^{1*}$\footnote{email: guangyou.qin@ccnu.edu.cn}
}

\address{$^{1}$ Key Laboratory of Quark and Lepton Physics (MOE) and Institute of Particle Physics, Central China Normal University, Wuhan 430079, China}

\address{$^{2}$ HiSilicon Research Department, Huawei Technologies Co., Ltd., Shenzhen 518000, China}

\date{\today}%

\begin{abstract}

The interactions of quarks and gluons are strong at non-perturbative region.
The equation of state (EoS) of a strongly-interacting quantum chromodynamics (QCD) medium 
can only be studied using the first-principle lattice QCD calculations.
However, the complicated QCD EoS can be reproduced using simple statistical formula
by treating the medium as a free parton gas whose fundamental degree of freedoms are 
dressed quarks and gluons called quasi-particles, with temperature-dependent masses.
We use deep neural network and auto differentiation to solve this variational problem in which
the masses of quasi gluons, up/down and strange quarks are three unknown functions, whose forms are represented by deep neural network.
We reproduce the QCD EoS using these machine learned quasi-particle masses,
and calculate the shear viscosity over entropy density ($\eta/s$) as a function of temperature of the hot QCD matter. 


\end{abstract}
\maketitle

\textit{Introduction} Given the masses of hundreds of different hadrons, one can compute the equation of state of non-interacting hadron resonance gas using simple statistical formulae~\cite{Peshier:1995ty,Levai:1997yx,Andronic:2012ut,Monnai:2021kgu}.
However, when the same procedure is applied to quarks and gluons whose interactions are described by quantum chromodynamics (QCD), the statistical formulae can not be performed easily.
One has to employ lattice QCD and sample field configurations using the Markov Chain Monte Carlo (MCMC) method or the normalizing flow method developed recently~\cite{Cossu:2017eys,Albergo:2019eim,Banuls:2019bmf,Caselle:2022acb}. 
In Ref.~\cite{Mroczek:2022oga}, the active learning method is used to exclude non-physical parameters in QCD equation of state (EoS).
Although the EoS of strongly interacting nuclear matter is difficult to calculate from the first principle lattice QCD calculations, it can be reproduced using simple statistical formulae, by assuming that the hot and dense quark gluon plasma (QGP) are made of non-interacting quasi-particles whose masses depend on the temperature of the medium~\cite{Goloviznin:1992ws,Brau:2009mp,Castorina:2011ja,Jakobus:2020nxw,1976A}.
Physically, these quasi-particles represent the in-medium quarks and gluons, each dressed with a large amount of neighboring partons.

Traditionally, one can provide a parameterized function for the temperature-dependent quasi-particle mass $M(T, \theta)$, where $\theta$ represents parameters that can be fixed using the lattice QCD data~\cite{HotQCD:2014kol,Liu:2021dpm}.
However, the extracted quasi-particle mass is sensitive to the prior function form with induced bias. It is also intractable to determine two or more unknown functions from the resulted EoS, using traditional methods. To solve this variational problem without introducing bias, we employ deep neural networks (DNN) to represent three unknown functions and solve it using auto-differentiation and optimization. 

DNN has achieved many unprecedented successes in solving inverse problems and variational problems in science because of its powerful representation capability~\cite{Gupta:2022vhe,Boehnlein:2021eym,thuerey2021pbdl}. This ability is supported theoretically by the universal approximation theorem ~\cite{Hornik1989MultilayerFN} using multi-layer feed-forward neural network, with sufficiently many hidden neurons. One successful application is to represent the solution of partial differential equations (PDEs) using DNN,
and train it by minimizing its violation to PDEs ~\cite{2018PINN,pde712178,khoo_lu_ying_2021,pde1528518,2017DGM,2018The,2021Deep,HAN2019108929,RAISSI2019686,Soma:2022vbb}. This technique relies on the auto-differentiation (auto-diff) of DNNs~\cite{JMLR:v18:17-468}. Different from classical numerical differentiation and symbolic differentiation, the auto-diff applies the chain rule and back-propagation to generate the output of the corresponding neuron and its gradient with respect to its inputs. Using auto-diff, different orders of derivatives of the DNN output with respect to its input are computed efficiently in analytical precision. Comparing with traditional methods, such as finite difference and finite element method, this method is mesh-free and overcomes the curse of dimensionality in high dimensional space. The same method can be generalized to represent physically unknown modules in a complex physical process~\cite{0Physics,Shi:2021qri} and optimize it using the final outcome.

Big data are required to train the neural network to rediscover some known physics. To be more data efficient, the neural network can be constructed in a way such that it satisfies some specific constraints automatically, e.g., to represent the solution of PDEs which satisfies initial and boundary conditions by construction \cite{pde712178,khoo_lu_ying_2021,pde1528518,2017DGM,2018The,2021Deep}, to represent the wave-function of many fermions that should be anti-symmetric \cite{HAN2019108929}, to represent magnetic fields whose divergence is zero by physical constraints \cite{Hendriks2020LinearlyCN}, or to represent the structure of molecules that is rotational equivalent \cite{e3nn}.




\begin{figure*}[htb]
\begin{centering}
\includegraphics[width=0.8\textwidth]{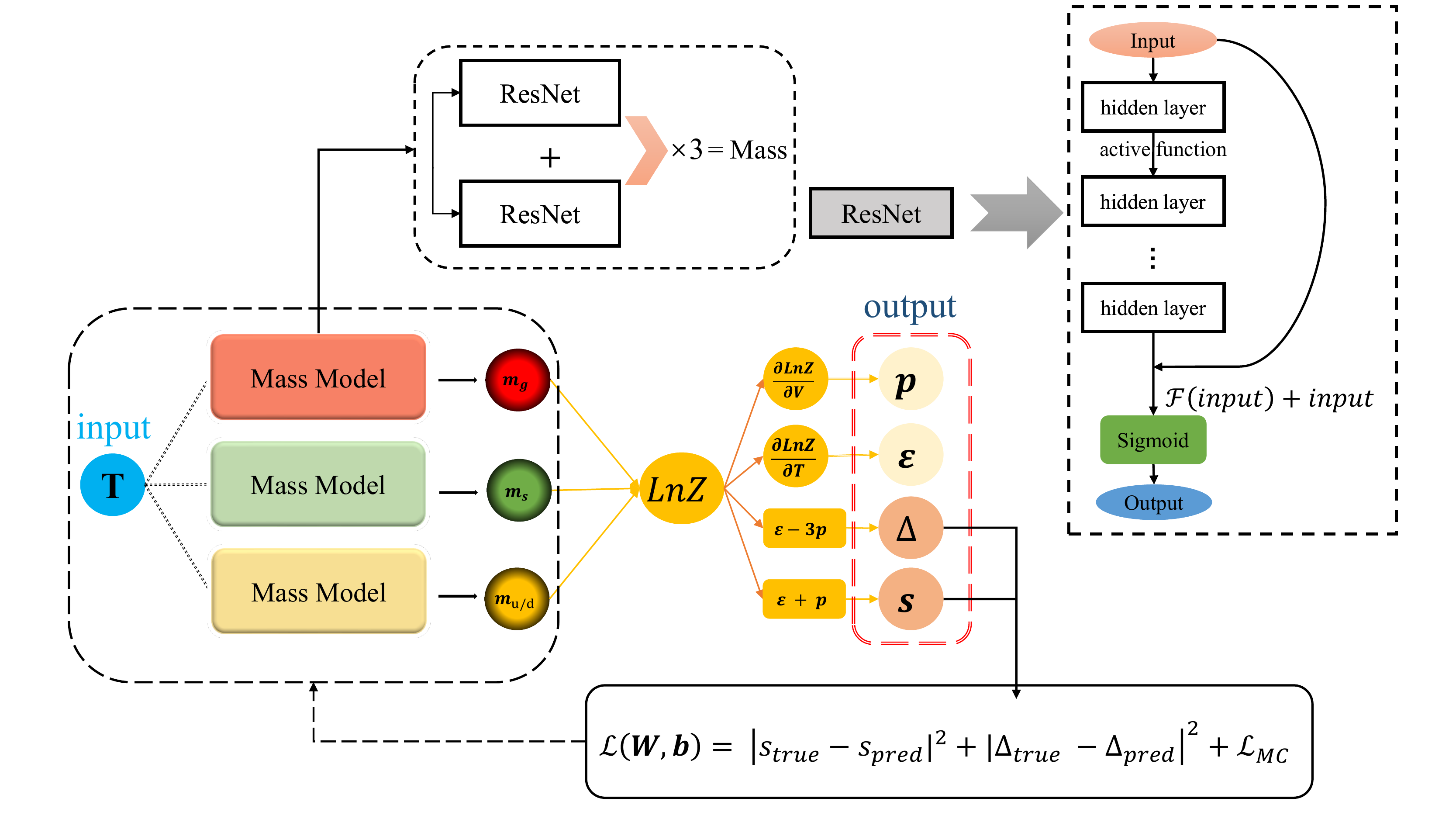}
\caption{(Color online) The framework of the neural networks for obtaining 3 mass functions in this study. Each mass model contains two residual neural networks. Each ResNet consists of 8 hidden layers with 32 neurons per layer. A swish-like activation function is used at the end of each hidden layer. }
\label{fig:network}
\end{centering}
\end{figure*}

\textit{Results and discussion} In this paper, we use three DNNs to represent the unknown mass of quasi quarks and gluons, as functions of local temperature of the quark-gluon plasma.
For an ideal gas of non-interacting quasi gluons and quarks, the partition function $Z$ is given by,
\begin{equation}
\ln Z(T) = \ln Z_g(T) + \sum_i \ln Z_{q_i}(T),
\label{eq:lnz}
\end{equation}
where $Z_g(T)$ is the contribution from gluons and $Z_{q_i}(T)$ is from quarks with flavor $i$.
They are computed from the momentum integration over the Bose-Einstein and Fermi-Dirac distributions, 
\begin{align}
\ln Z_g(T) =& - \frac{16 V}{2 \pi^{2}} \int_{0}^{\infty}p^{2} dp 
\nonumber\\
& \ln \left[ 1 - \exp \left(-{1 \over T}\sqrt{p^{2}+m_g^{2}(T)}\right) \right], \\
\ln Z_{q_i}(T) =& + \frac{12 V}{2 \pi^{2}} \int_{0}^{\infty}p^{2} dp
\nonumber\\
&  \ln \left[ 1 + \exp \left(- {1 \over T}\sqrt{p^{2}+m_{q_i}^{2}(T)}\right) \right],
\label{eq:lnz_gluon}
\end{align}
where $V$ is the volume of the system, $p$ is the momentum of quasi-particles, $T$ is the temperature,
$m_g(T)$ and $m_{q_i}(T)$ are the temperature-dependent masses for gluons and quarks, respectively.
Note that the degree of freedom (DOF) for gluons is $16 = 2 \times 8$ in SU(3) gauge theory with 2 polarizations for each of the 8 gluons, the DOF for quark flavor $q_i$ is $12=2\times2\times 3 = N_a \times N_s \times N_c$, where $N_a=2$ counts the DOF of quark and anti-quark, $N_s = 2$ for spin degeneracy and $N_c=3$ for number of colors.  

For a system of quarks and gluons in thermal equilibrium, with temperature high enough
to excite strange and anti-strange quark pairs $s\bar{s}$ from the vacuum, 
the partition function $Z(T)$ is given by,
\begin{align}
\ln Z(T) = \ln Z_g(T) + \ln Z_{u,d}(T) +  \ln Z_s(T),
\label{eq:lnz_total}
\end{align}
which depends on the temperature-dependent masses of gluons, and up, down and strange quarks.
In practice, we use the same temperature-dependent mass for up and down quarks.
As a result, there will be 3 temperature-dependent mass functions, $m_{u/d}(T, \theta_1)$ for up and down quarks, $m_{s}(T, \theta_2)$ for strange quark and $m_{g}(T, \theta_3)$  for gluons, where $\theta_1$, $\theta_2$ and $\theta_3$ are the parameters in DNN shown in Fig.~\ref{fig:network}.

The resulting pressure and energy density are computed using the following statistical formulae,
\begin{align}
P(T) &= T \left( \frac{\partial \ln Z(T)}{\partial V} \right)_{T}, \\
\epsilon(T) &=\frac{T^{2}}{V}\left(\frac{\partial \ln Z(T)}{\partial T}\right)_{V},
\label{eq:eos}
\end{align}
Once the energy density and pressure are provided by the deep neural network,
the entropy density $s(T)=\frac{\varepsilon + P}{T}$
and trace anomaly $\Delta(T) =  {(\varepsilon - 3P)/ T^4}$ as a function of temperature can be calculated accordingly.
The main goal of this work is to determine $m_{u/d}(T, \theta_1)$, $m_{s}(T, \theta_2)$ and $m_{g}(T, \theta_3)$ by minimizing the mean-square-error between network prediction and the Lattice QCD calculations of $s(T)$ and $\Delta(T)$ .
In the forward process, these 3 mass functions represented by DNNs are used in the Fermi-Dirac distribution functions for quarks
and Bose-Einstein distribution functions for gluons to compute the momentum integration in the partition function, as shown in Eq.~(\ref{eq:lnz_gluon}).
The integration is implemented numerically with Gauss quadrature method using operators provided by
tensorflow, in which the derivatives of partition function shown in Eq.~(\ref{eq:eos}) 
are provided automatically by auto-diff, which has analytical precision in principle.

\begin{figure}[htp]
\centering
\includegraphics[width=0.48\textwidth]{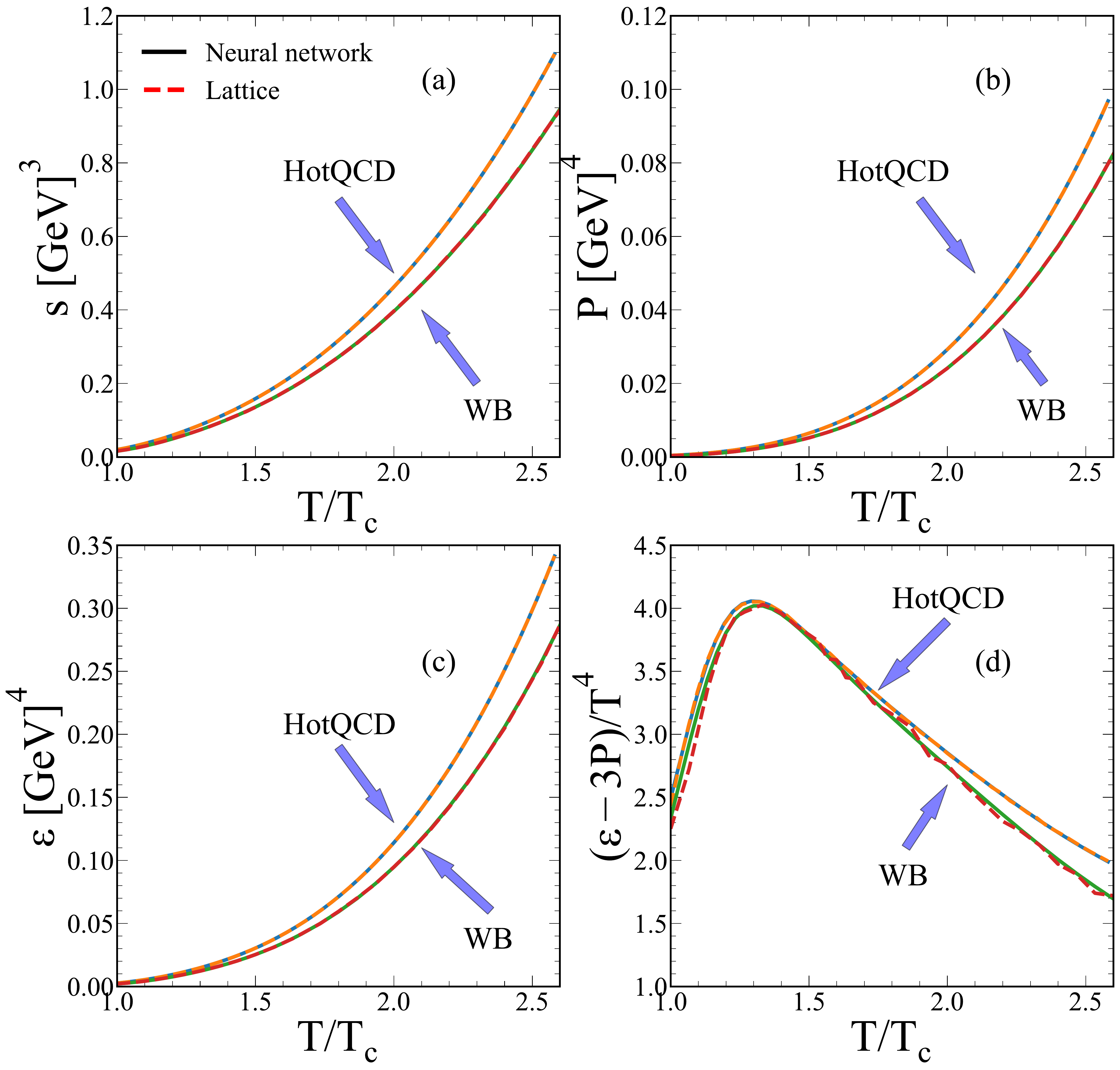}
\caption {\label{fig:MC_HotQCD}(Color online) Training and prediction results within mass constraints from HotQCD and WB lattice QCD (WB), the $T_c$ of the former is 0.155 GeV and the latter one is 0.150 GeV. (a), (b), (c) and (d) indicate entropy density $s$, pressure $P$, energy density $\epsilon$ and trace anomaly $\Delta$, respectively,  as functions of temperature. The solid lines are the DNN predictions and the dashed lines are the lattice QCD data. The red and indigo lines represent the HotQCD group; the green and red lines refer to the WB group.}
\end{figure}

\begin{figure*}[htp]
\centering
\includegraphics[width=0.8\textwidth]{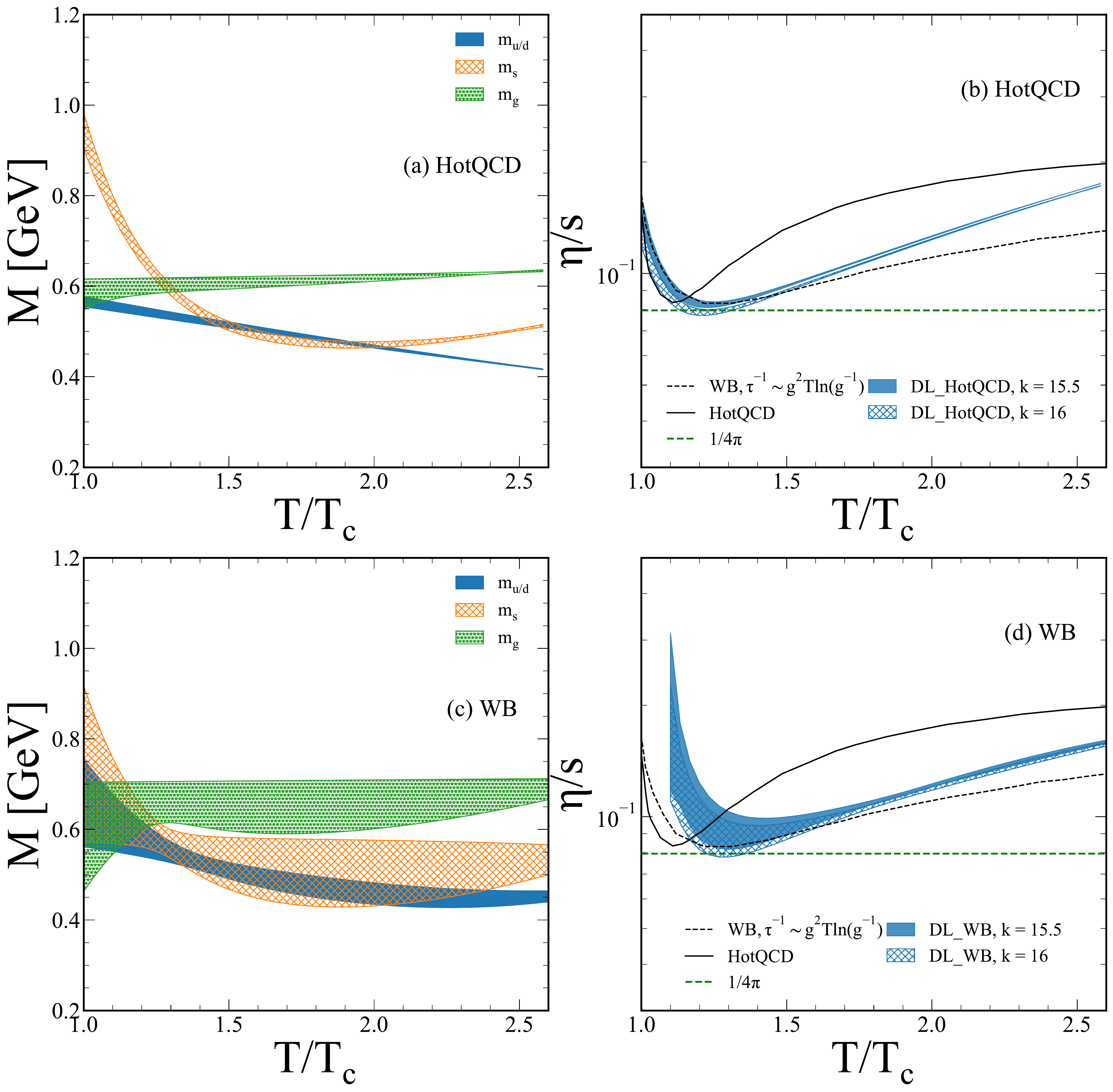}
\caption {\label{fig:MC_WB}(Color online) Masses and $\eta/s$ given by DNN within mass constraints from HotQCD and WB lattice QCD: the upper row is HotQCD and the bottom row is WB lattice QCD. Panels (a) and (c) are presented as a function of temperature for different masses: the blue bands are $u/d$ quarks, the orange one is $s$ quark and the green one is gluon. Panels (b) and (d) correspond to the $\eta/s$ calculated using masses given by (a) and (c). The blue shaded bands and the hatched region are results calculated using k with different values, 15.5 and 16 respectively. The black solid and dotted curves are results from Ref.~\cite{Plumari:2011mk}.}
\end{figure*}

Before training, the neural network parameters $\theta_1,\theta_2$ and $\theta_3$ are initialized with random numbers. As a result, the partition function as well as the calculated entropy density and trace anomaly deviate from the lattice QCD predictions. After trained using stochastic gradient descent algorithm (Adam), the learned mass functions reproduce not only $s$ and $\Delta$, but also $\epsilon$ and $P$, as shown in 
Fig.~\ref{fig:MC_HotQCD}.
The solid lines shown in Fig.~\ref{fig:MC_HotQCD} are computed using the DNN-learned mass functions of quasi quarks and gluons. The dashed lines are provided by Lattice QCD. Using the data from HotQCD and WB lattice QCD groups, two groups of mass functions are learned separately and can describe data. DNN shows its powerful representation ability that might be used in various variational problems.


Using these temperature-dependent quasi-particle masses, we compute the shear and bulk viscosity of the hot and dense QCD matter using the linear response theory~\cite{Xu:2007ns,Greco:2008fs,Wesp:2011yy,Policastro:2001yc,Kovtun:2004de}.
The left column (a) and (c) in 
Fig.~\ref{fig:MC_WB} show the learned quasi particle masses as a function of temperature, using HotQCD data and WB Lattice QCD data respectively. The qualitative trends of these learned mass functions are similar for two groups of lattice QCD data. The dressed masses of $u/d$ and $s$ quarks decrease as the temperature increases. The dressed mass of gluons increases slowly as the temperature increases. At high temperature, the separations between different mass functions are constrained using analytical solutions from perturbative QCD.
The uncertainty bands are provided by training the neural network multiple times.  


The right columns (b) and (d) show the corresponding $\eta/s$ as a function of $T$ using the learned mass functions (blue bands), compared with the WB and HotQCD results using parameterized mass functions together with the $1/(4\pi)$ limit from Ads/CFT calculations. Using two values for the cutoff parameter $k$ (15.5 and 16), the tendency of $\eta/s(T)$ are similar to the results using parameterized mass functions. 
All curves of $\eta/s(T)$ first drop to a minimum and then rise again as the temperature increases. Specifically, the $\eta/s(T)$ obtained using DNN method exists a minimum value at $T \approx 1.25~T_c$. At high temperature region, the value of $\eta/s$ is in the range 0.1 and 0.2, which is consistent with the predictions of heavy ion collisions (HIC)~\cite{Romatschke:2007mq,Schenke:2010rr,Heinz:2013th}. Different from the approach of Ref.~\cite{Plumari:2011mk}, which requires the coupling $g$ by fitting the lattice data, we can directly extract mass functions using DNN and then calculating the coupling $g$. Note that when $k$ is 15.5, $\eta/s$ will have a negative value at low temperature which are truncated in the figure. 

\begin{figure}[htp]
\centering
\includegraphics[width=0.48\textwidth]{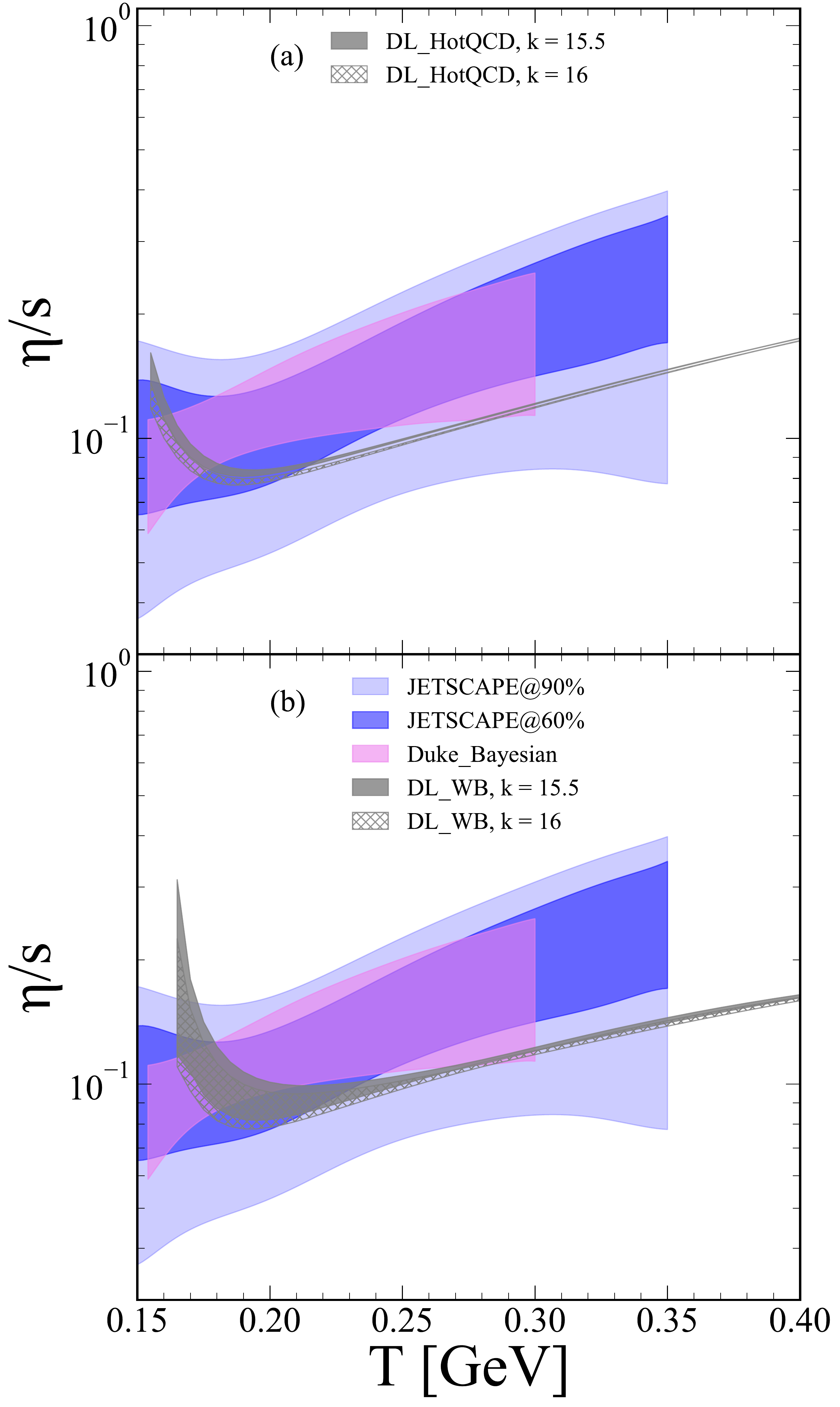}
\caption {\label{fig:nsbaye}(Color online) The $\eta/s(T)$ calculated using DNN as compared with Bayesian analysis. The $\eta/s(T)$ from DNN are shown in grey band and hatched band using training data from (a) HotQCD and (b) WB lattice QCD. The hatched band is calculated using $k=16$, and the grey band uses $k=15.5$. The light blue and dark blue shaded areas are the 90\% and 60\% confidence intervals from the Ref.~\cite{JETSCAPE:2020avt}. The pink band are 90\% credible region for the $\eta/s$ from the Duck Beyasian analysis Ref.~\cite{Bernhard:2019bmu}.}
\end{figure}

Fig.~\ref{fig:nsbaye} compares the $\eta/s(T)$ from our calculations with two groups of Bayesian analysis~\cite{Bernhard:2019bmu,JETSCAPE:2020avt}. In Bayesian analysis, parameterized $\eta/s(T)$ is used in relativistic hydrodynamics to compute the unnormalized posterior distribution of the parameters in $\eta/s(T)$ given experimental data, which can be sampled using Markov Chain Monte Carlo (MCMC) method by random walk in the parameter space. The sampled $\eta/s(T)$ thus have uncertainty bands inherited from the error bars of experimental data. Our $\eta/s(T)$ using both WB and Lattice QCD EoS are     basically consistent with the results of JETSCAPE within 90\% confidence area especially in $T>0.17$~GeV. The location of the minimum of $\eta/s(T)$ agrees with JETSCAPE within 60\% confidence level. This is different from the Duke Bayesian where the minimum appears at a lower temperature than JETSCAPE. 
Note that Duke Bayesian parameterizes $\eta/s(T)$ in such a way that the location of the minimum is fixed to $T_c$ and $\eta/s(T)$ increases monotonically above $T_c$~\cite{Bernhard:2019bmu}. It is well known that the posterior distribution in Bayesian analysis is affected by the a priori distribution. The variational function form of $\eta/s(T)$ using DNN does not have such constraints.


\textit{Discussion} The problem we solved in the present paper is an inverse problem of statistical mechanics.
There are three mass functions whose forms are unknown.
Different function forms will lead to different partition functions and equation of states.
The task is to look for one group of variational mass functions that will
lead to the correct partition function and the equation of state of strongly coupled nuclear matter.
Before doing the calculation, we do not know whether different combinations will lead to the same result. 

The DNN can map the relationship between temperature $T$ and mass $M$ by minimizing the loss function $\mathcal{L}$. Being trained with entropy density $s$ and trace anomaly $\Delta$, other thermodynamic quantities such as pressure $P$ and energy density $\epsilon$ are in good agreement with the result of HotQCD and WB lattice QCD. In addition, we also note that the mass constraints at high temperatures are also well captured by the DNN. We show in the supplementary material that using the upper bound and lower bound of the lattice QCD EoS, we can also provide an uncertainty band for the extracted mass. However, since different data points have correlations, we do not treat this as a serious uncertainty analysis.

Using DNN to solve the complex partial differential equations always encounter the problem of convergence. As the network parameters are initialized using random numbers, different initial locations in the parameter space may lead to different solutions. We notice that training the neural network with different random seeds sometimes lead to worse solutions than our current result. In practice, we only pick solutions with small loss and treat different $m_{u/d}(T)$, $m_s(T)$ and $m_g(T)$ as uncertainties from neural network.
We find that adding training data, especially at low temperature, can improve the efficiency of DNN in solving mass functions. 
Using physical constraints at high temperature also helps the network to converge.


Parton cascade programs are widely used to simulate high energy heavy ion collisions.
However, traditional parton cascade failed to reproduce the QCD equation of state
extracted from first principle lattice QCD calculations~\cite{Zhang:1997ej,Xu:2004mz}.
Using the quasi partons learned by the network, 
non-perturbative interactions between quarks and gluons are encoded in the temperature dependent mass.
It is thus be possible to construct a quasi parton cascade (QPC) program that
can reproduce the lattice QCD EoS with elastic scatterings only~\cite{Xu:2007ns,Greco:2008fs,Wesp:2011yy}.


\textit{Method} 
DNNs are used to represent the temperature dependent quasi-particle mass $m_{u/d}(T)$, $m_s(T)$ and $m_g(T)$. Using statistical formulae, 
we compute EoS of strongly coupled nuclear matter, and then compare it with lattice QCD EoS.
The root mean square error (RMSE) is a functional of 3 unknown mass functions; 
minimizing the RMSE will help to find them automatically.
After being trained, the network returns 3 learned functions.

Once the masses become temperature-dependent, one may introduce the temperature-dependent bag constant $B(T)$~\cite{Carlson:1982er,Gorenstein:1995vm,Dorokhov:1983uj,Hansson:1985hm}. According to Eq.~(\ref{eq:eos}),
\begin{align}
  \epsilon &= \sum_i\frac{d_i}{2\pi^2} \Bigg(\int_{0}^{\infty}p^2 dp\frac{\sqrt{p^2 
   + m^2}}{\exp(\frac{\sqrt{p^2+m2}}{T}) \mp 1} 
   \nonumber\\
   &+  \int_{0}^{\infty}\frac{p^2dp^2}{\sqrt{p^2+m^2}} \frac{m\frac{\partial m}{\partial T}}{\exp(\frac{\sqrt{p^2+m^2}}{T}) \mp 1 }\Bigg)
   \nonumber\\
   &= \varepsilon_{id}(T) + B(T), 
  \label{eq:bag_func}
\end{align}
where the $d_i$ is the DOF of particles $i$. In order to keep the statistical mechanics and thermodynamics consistent $\varepsilon + p = sT$, the pressure is rewritten as follows,
\begin{align}
  P(T) \rightarrow P_{id}(T) - B(T), 
  \label{eq:bag}
\end{align}
where $\varepsilon_{id}(T)$ and $P_{id}(T)$ represent the ideal gas formulas for particles of given mass. Thus $B(T)$ can be conveniently calculated once the quasi-particle mass is obtained. We determine the temperature-dependent masses of quarks and gluons by fitting $s(T)$ and $\Delta(T)$ with the lattice QCD result.
The pressure and energy density provide additional data for testing.

To further verify other thermodynamic properties, we also calculate the shear viscosity to entropy density ratio $\eta / s$, a vital transport coefficient of QGP~\cite{Majumder:2007zh,Romatschke:2007mq,Demir:2008tr,Koide:2009sy,Song:2010mg,Bernhard:2019bmu}.
 There are many analytical methods, such as Green-Kubo method, Chapman-Enskog and and relaxation-time approximations, etc.~\cite{Chen:2007xe,Plumari:2011mk,Plumari:2012ep}. 
 In this work, we use the following expression~\cite{Plumari:2011mk}:
\begin{align}
\eta = \frac{1}{15T}\sum_i d_i\int\frac{d^3p}{(2\pi)^3}\tau_i\frac{{p}^4}{E_a^2}f_i(1\mp f_i),
\label{eq:eta}
\end{align}
where $\tau_i$ is collisional relaxation time for the different particles and $f_i$ is the equilibrium distribution function of particles $i$. The collisional relaxation time $\tau_i$ for quarks and gluons can be expressed as follows~\cite{Peshier:2004bv,Peshier:2005pp,Khvorostukhin:2010aj}:
\begin{align}
\tau_q^{-1} = 2 \frac{N_C^2 - 1}{2N_C} \frac{g^2T}{8\pi}\ln\frac{2k}{g^2},\tau_g^{-1} = 2N_C\frac{g^2T}{8\pi}\ln\frac{2k}{g^2},
  \label{eq:omega}
\end{align}
where $N_f=3$ is the number of parton flavors considered here. 
Note that $k$ is the only parameter to be tuned. As we will see, it influences the minimum of $\eta/s$. 
If $2k\leq g^2$, the value of $\tau^{-1}$ will be negative and vice versa. 
The coupling $g^2$ can be derived from the effective masses in a perturbative approach~\cite{Khvorostukhin:2010aj}:
\begin{align}
m_g^2 = \frac{1}{6}g^2(N_C + \frac{1}{2}N_f)T^2,m_{u/d}^2 = \frac{N_C^2 - 1}{8N_C}g^2T^2.
\label{eq:mg}
\end{align}
Note that in traditional approach, $g^2$ is usually parameterized and fixed by fitting to the lattice QCD data. 
In our approach, since the effective masses can be obtained from DNN, there is no need to parameterize $g^2$. 
Instead, it can be directly derived from the Eq.~(\ref{eq:mg}) as follows:
\begin{align}
g^2 = \frac{m_g^2 + m_{u/d}^2}{\bigg[ \frac{1}{6}(N_C + \frac{1}{2}N_f) + \frac{N_C^2 - 1}{8N_C} \bigg]T^2},
\label{eq:g2}
\end{align}
Substituting Eqs. (\ref{eq:omega}) and (\ref{eq:g2}) into the Eq.~(\ref{eq:eta}), we can calculate $\eta$ conveniently.

Figure~\ref{fig:network} shows the framework of our study.
Three developed residual neural networks are used as the mass models for $m_{u/d}(T, \theta_1)$, $m_{s}(T, \theta_2)$ and $m_{g}(T, \theta_3)$. 
Each mass model represents a residual neural networks (ResNet), which have 8 hidden layers with 32 neurons and a swish-like activation function. The sigmoid activation function is used at the end of the each ResNet and then added to the outputs of two ResNets. 
Finally, the output of these mass models are fed to an integration layer using 25 points Gaussian quadrature to compute the partition function $\ln Z(T)$ numerically.
At this step, the auto-diff tools provided by tensorflow library is used to compute the derivatives of $\ln Z(T)$ with respect to the input temperature $T$ of the neural network.
The resulted EoS and the Lattice QCD data are used as the training objectives and test objectives.

The training objective is to minimize the loss which contains two parts.
One is the RMSE of the entropy density and trace anomaly between the network outputs and lattice QCD calculations.
The other is the physical constraint. For example, the pQCD calculation~\cite{Levai:1997yx} shows a mass hierarchy $m_g > m_{u/d},m_s$ at $T>2.5T_{cut}$, where $T_{cut}$ is taken as 0.150 GeV in this study. 
To implement this physical constraint, we first note the ratio of the gluon to quark effective masses at high temperature,
\begin{align}
R_{g/q} =  \frac{M_{g,T>2.5T_{cut}}}{M_{q,T>2.5T_{cut}}} = \sqrt{\frac{3}{2}(\frac{N_C}{3} + \frac{N_f}{6})}.
\label{eq:MCC}
\end{align} 
The constraint between $m_{u/d}$ and $m_g$ can be written as:
\begin{align}
\mathcal{L}_1 = \left| R_{g/q} - \frac{3}{2} \right|.
\label{eq:LM1}
\end{align}
For $s$ quark, the constraint at high temperature is set to
\begin{align}
\mathcal{L}_2 = \left| \frac{m_s - m_{u/d}}{\overline{m}_s - \overline{m}_{u/d}} -1\right|.
\label{eq:LM2}
\end{align}
where $\overline{m}_s \approx 95$~MeV and $ \overline{m}_{u/d} \approx 5$~MeV are current masses of quarks. 
The total loss function is then:
\begin{align}
\mathcal{L} = (s_{true}-s_{pred})^2 +  (\Delta_{true}-\Delta_{pred})^2 + \mathcal{L}_{MC}.
\label{eq:LMCC}
\end{align}
where the constraints $\mathcal{L}_{MC} = (\beta_1 \mathcal{L}_1 + \beta_2 \mathcal{L}_2)^2$ can be understood as the regularization of masses. To minimize the influence of mass constraints on the prediction of entropy density, the values of $\beta_1$ and $\beta_2$ should not be too large. 
In this work, we take $\beta_1=0.01$ and $\beta_2=0.0006$.


\begin{acknowledgments}
This work was supported in part by the NSFC under grant Nos. 12075098, 12225503, 11890710, 11890711, 11935007.
LG Pang and FP Li also acknowledge the support provided by Huawei Technologies Co., Ltd.
We gratefully acknowledge the extensive computing resources provided by the Nuclear Computing Center of Central China Normal University. The contributions of Dr. Hong-Liang L\"u are non-Huawei achievements.
\end{acknowledgments}

\bibliography{ref}

\begin{appendix}

\section{Without physical prior}
\begin{figure}[htp]
\centering
\includegraphics[width=0.48\textwidth]{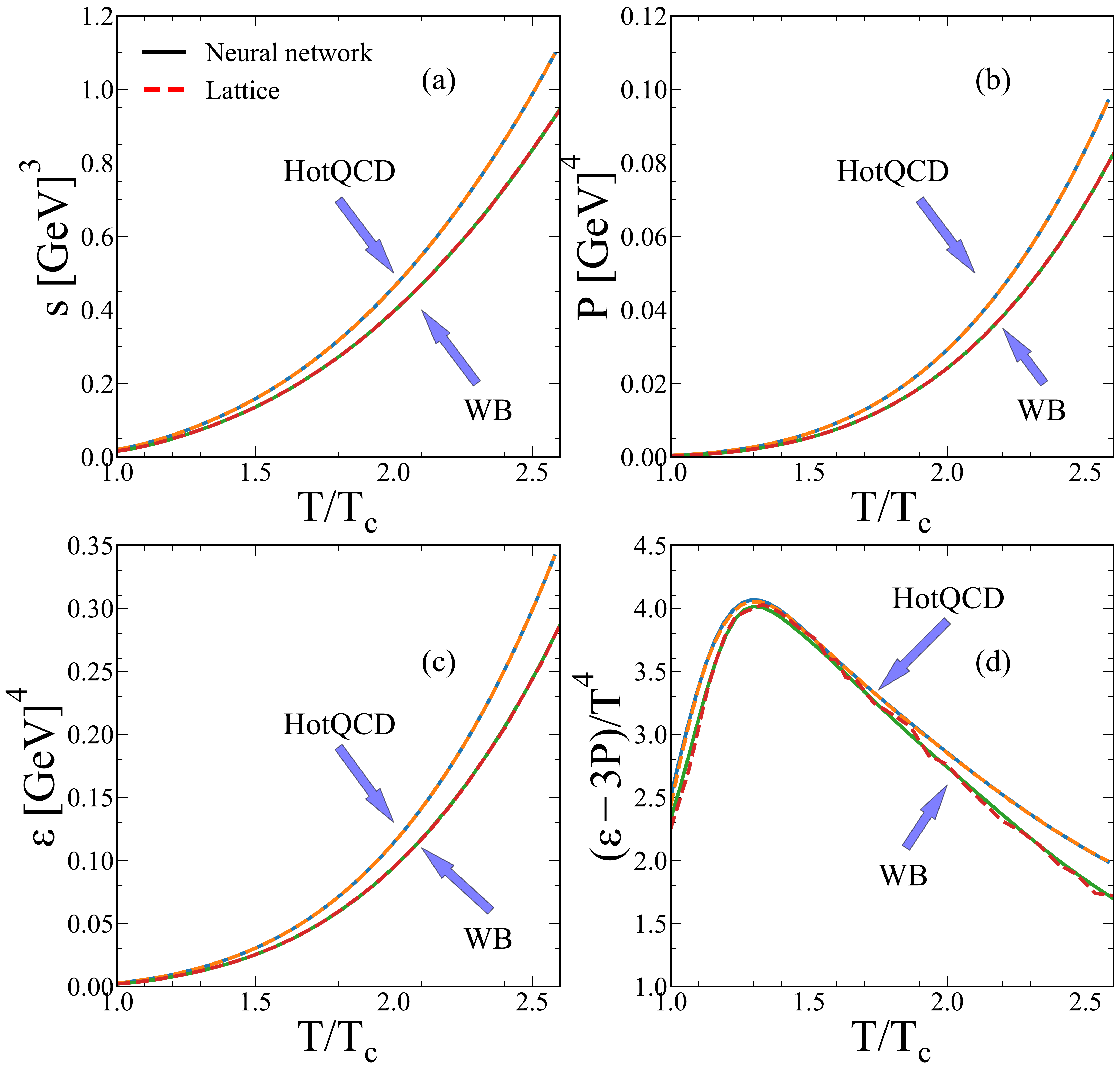}
\caption {\label{fig:woMC}(Color online) Training and prediction results within mass constraints from HotQCD and WB lattice QCD (WB). (a), (b), (c) and (d) respectively indicate entropy density $s$, pressure $P$, energy density $\epsilon$ and trace anomaly $\Delta$ as functions of temperature. The solid line is the DNN prediction data and the dashed line is the lattice QCD data. The orange and blue line represent HotQCD. And the green and red line refer to the WB lattice QCD.}
\end{figure}

Shown in Fig.~\ref{fig:woMC} (a) is the entropy density $s(T)$ given by deep neural network (DNN)
as compared with the HotQCD and WB lattice QCD theoretical calculations. 
We selected 50 points from two lattice group theoretical calculations as input data for training. 
Two curves match each other perfectly. 
This indicates that the entropy density sector of the EoS of strongly interacting nuclear matter
can be readily reproduced using non-interacting quasi particles.
Our DNN has enough representation power to learn 
these 3 unknown mass functions to reproduce Lattice QCD EoS.

As the entropy density contains both contributions from energy density and pressure,
one would guess whether there is compensation between these two quantities.
Shown in Fig.~\ref{fig:woMC} (b, c) are the pressure $P(T)$ and energy density $\epsilon (T)$ 
computed from the learned 3 mass functions by the neural network,
as compared to theoretical calculations assuming zero bag constant.
These two thermodynamic quantities match the theoretical calculations as well, 
though only entropy densities at different temperatures are used for training.

Fig.~\ref{fig:woMC} (d) shows the trace anomaly from DNN compared with Lattice results.
During training, we notice that the trace anomaly converges slowest as compared
with entropy density, energy density and pressure.
Usually it takes much longer to converge to the correct trace anomaly 
as given by Lattice QCD calculations.
Sometimes the final trace anomaly still deviates from theoretical calculations
long after the entropy density curve converges to ground truth.
The parameters of the network are initialized randomly for each training,
as a result, it is not guaranteed to converge to the theoretical trace anomaly 
even if the entropy density curve is reproduced.
Since we can always treat trace anomaly as training data instead of testing data,
the convergence of trace anomaly during test is not a big concern. In addition, we can make the trace anomaly converges better with more training points. Naturally, we can also obtain perfect solutions by retraining the DNN.

\begin{figure*}[htp]
\begin{centering}
\includegraphics[width=0.98\textwidth]{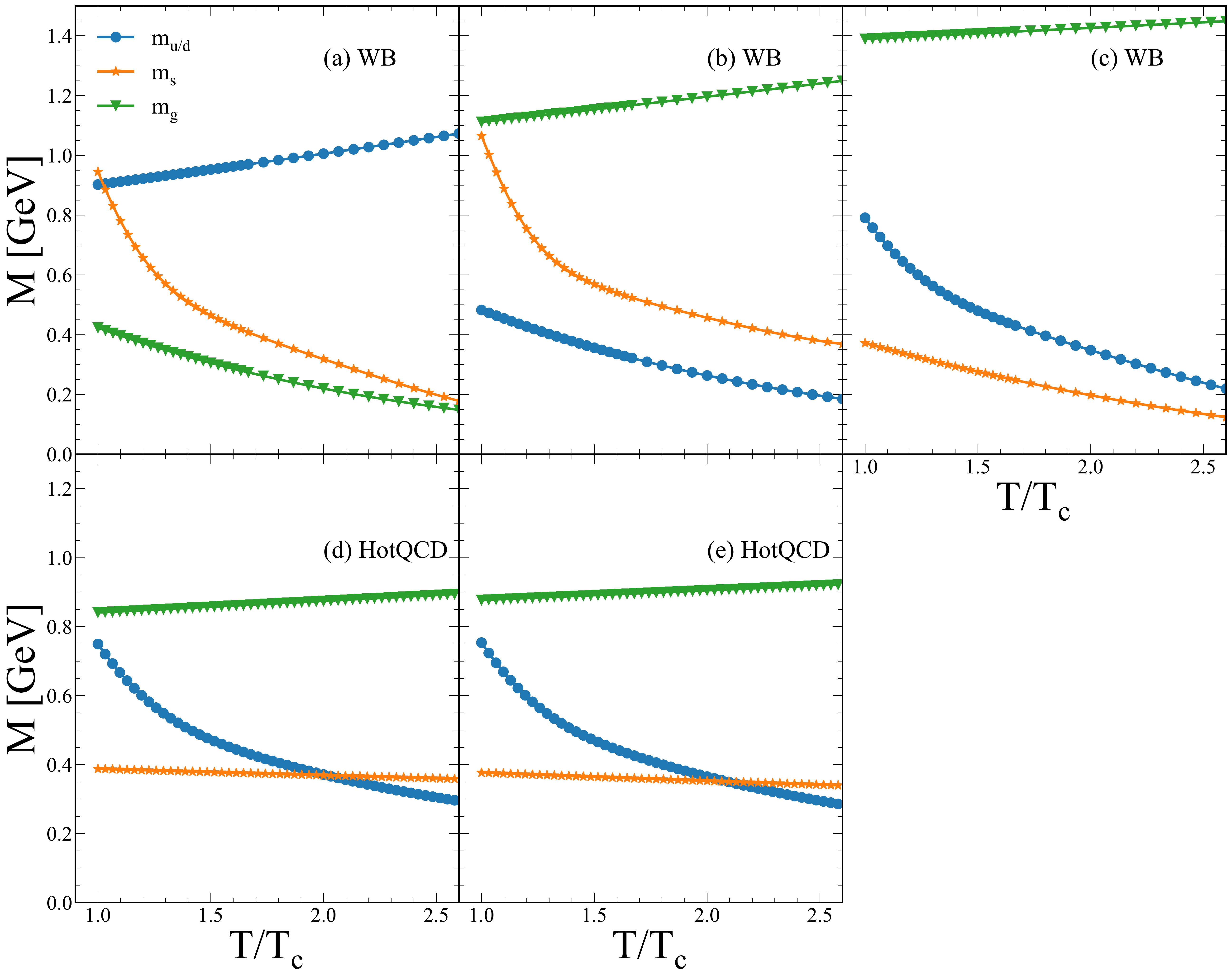}
\caption{(Color online) Masses as a function of temperature $T$ from two lattice groups, the upper panels are WB lattice QCD and the lower are HotQCD. The blue curve is $m_{u/d}$, the green one is $m_s$ and the orange one is $m_g$. Note the mass constraint are not used here. }
\label{fig:woMass}
\end{centering}
\end{figure*}

Shown in Fig.~\ref{fig:woMass} are the three temperature dependent mass functions 
learned by the DNN without the mass constraint. Using the entropy density $s$ and trace anomaly $\Delta$ as the constraint during training leads to the non-uniqueness of the solution.
The result indicates that the mass obtained from DNN is not unique without the mass constraint.
These functions are not exactly the same as what has been derived using
other methods, such as two-parameter fitting functions or Bayesian method~\cite{Liu:2021dpm}.
Since the quasi particle model is an effective theory, 
we will not worry this difference as long as they can reproduce
the EoS of strongly interacting nuclear matter.
It also tells that there are more than one ways to get the same QCD EoS 
using different temperature dependent masses for quasi quarks and gluons.


\section{Distribution function}
 Fig.~\ref{fig:disf} shows that the distribution function $f(p)$ of HotQCD and WB lattice QCD for three temperatures 0.2, 0.3 and 0.4 GeV. One can see that the $f(p)$ with mass constraint follows the order of $m_{u/d}$, $m_s$ and $m_g$ in high momentum.
 
\begin{figure*}[htp]
\centering
\includegraphics[width=0.98\textwidth]{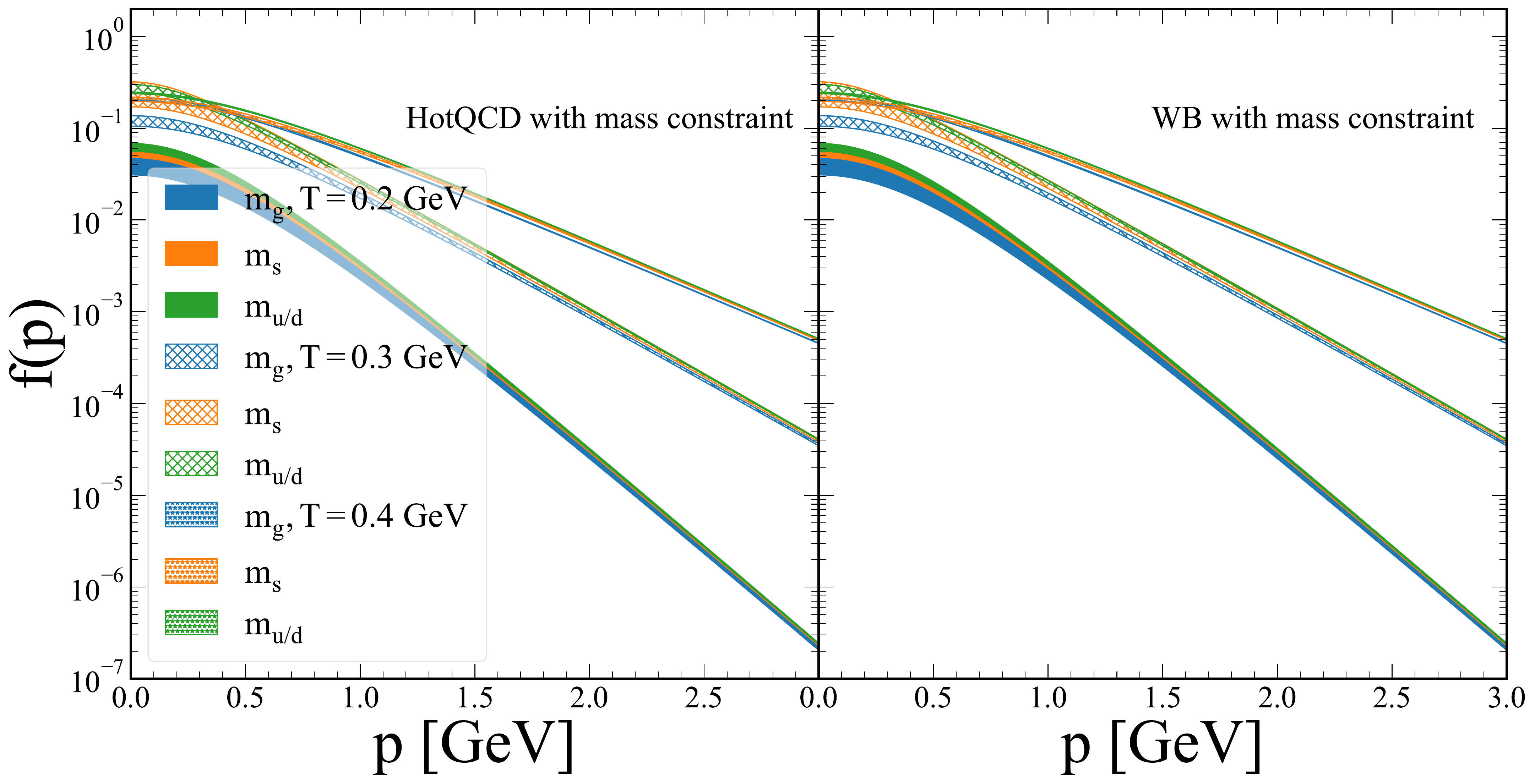}
\caption {\label{fig:disf}(Color online) The distribution function of HotQCD and WB lattice QCD at three temperatures 0.2, 0.3 and 0.4 GeV. The blue shaded is $m_g$, the orange shaded is $m_s$ and the green one is $m_{u/d}$.}
\end{figure*}

\section{The influence of $\beta_2$}

\begin{figure}[htp]
\begin{centering}
\includegraphics[width=0.48\textwidth]{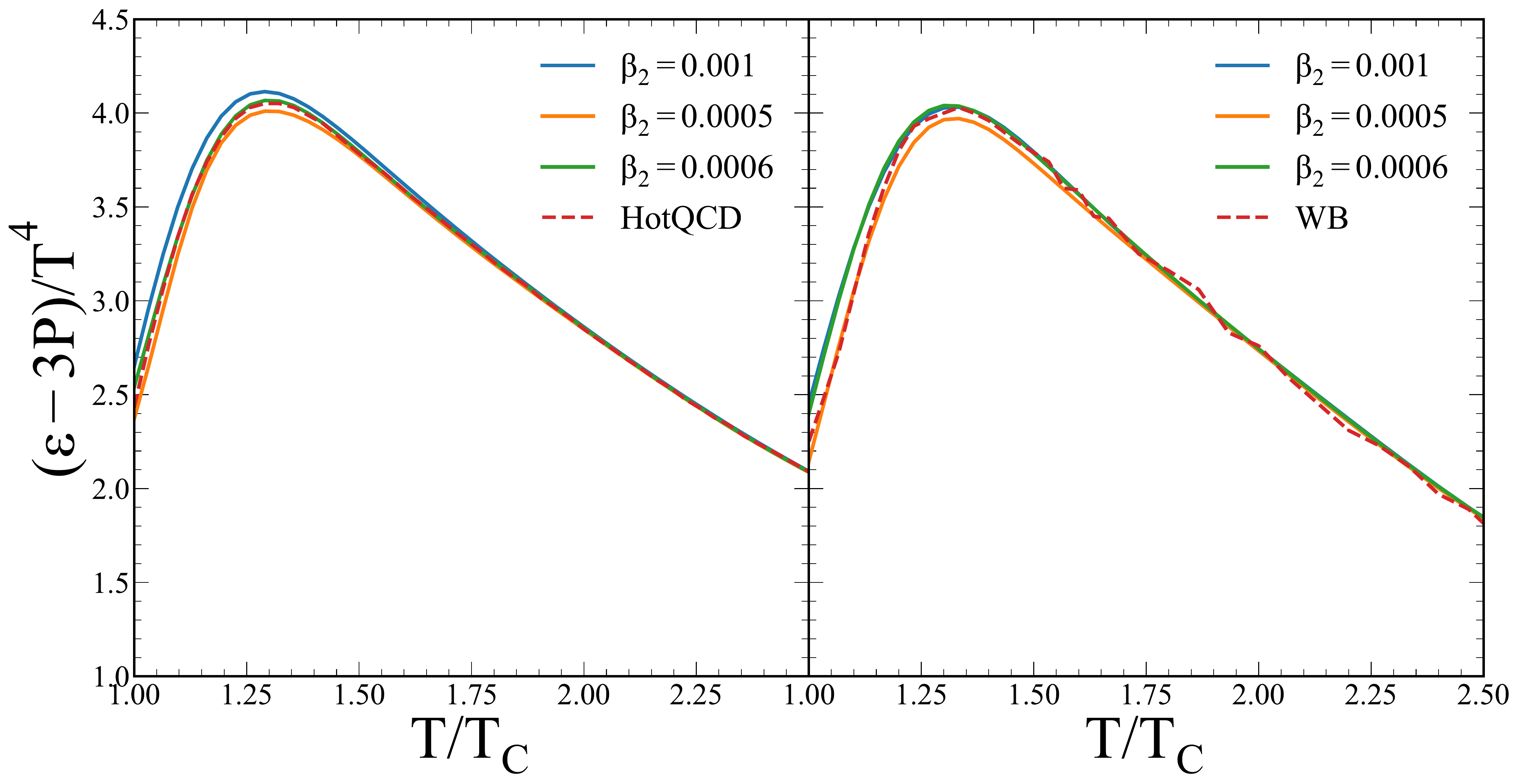}
\caption{(Color online). The trace anomaly with different $\beta_2$ values and the same $\beta_1 = 0.01$ as predicted by DNN. The dash line is the result of lattice QCD, and the solid is the DNN result. }
\label{fig:beta}
\end{centering}
\end{figure}

In Eq.~(\ref{eq:LMCC}), we introduce the regularization parameter $\beta_1$ and $\beta_2$ to adjust the influence of mass constraints. Fig. \ref{fig:beta} demonstrates that the different $\beta_2$ will change the consequent of trace anomaly. If $\beta_2$ is too large, the $\epsilon - 3P$ will be larger at low temperature, which may be related to the temperature range of the mass constraints. 

\section{Bag constant B(T)}

\begin{figure}[htp]
\begin{centering}
\includegraphics[width=0.48\textwidth]{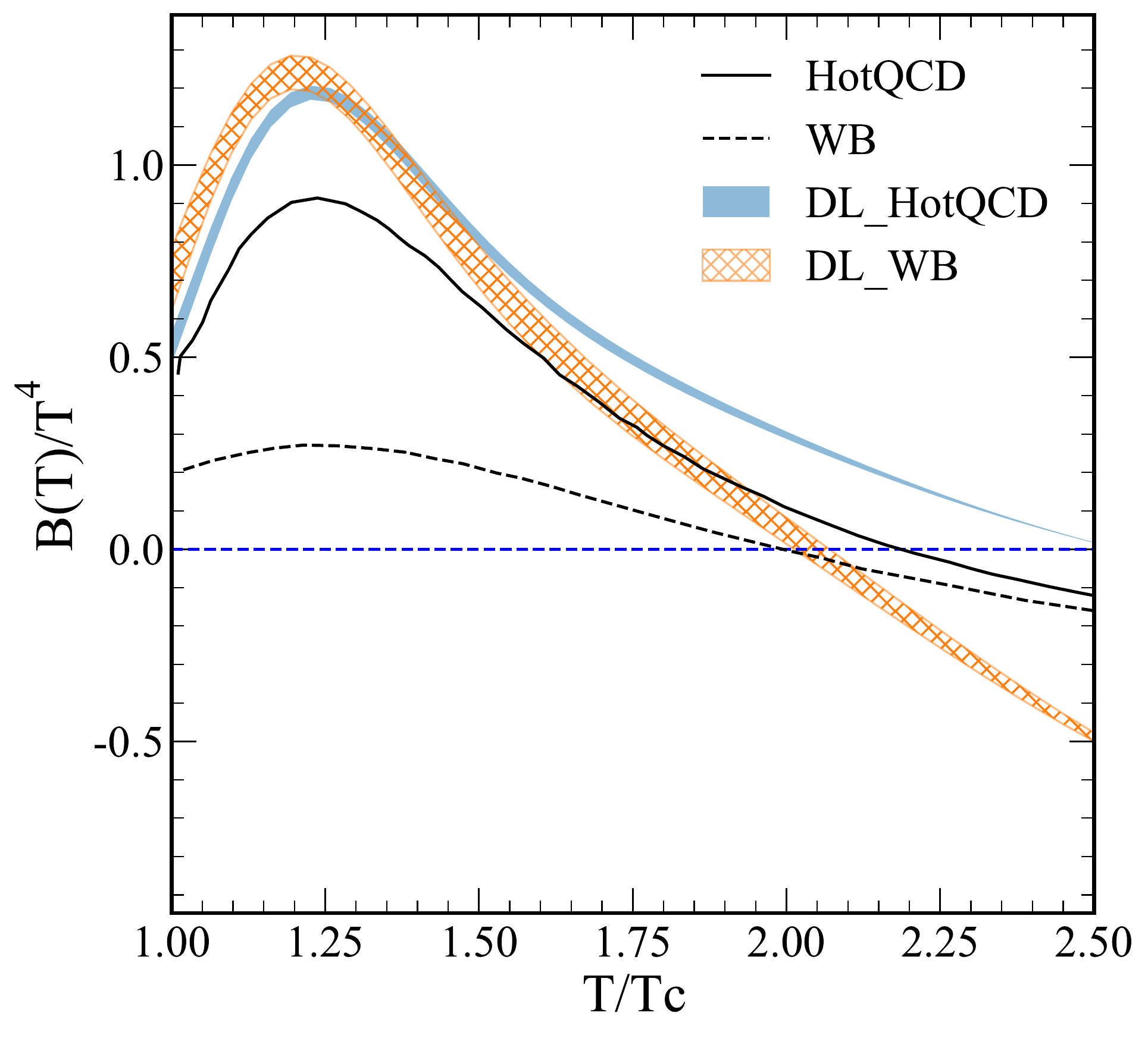}
 \caption{(Color online) Temperature-dependent bag constant functions.  }
\label{fig:bag}
\end{centering}
\end{figure}

Fig.~\ref{fig:bag} shows that the temperature-dependent bag constant functions comparing with other quasi-particle models. The orange and blue band are the results of our method using Eq.~(\ref{eq:bag_func}). The dash and solid line are results from Ref.~\cite{Plumari:2011mk}. 

\begin{figure*}[htp]
\begin{centering}
\includegraphics[width=0.98\textwidth]{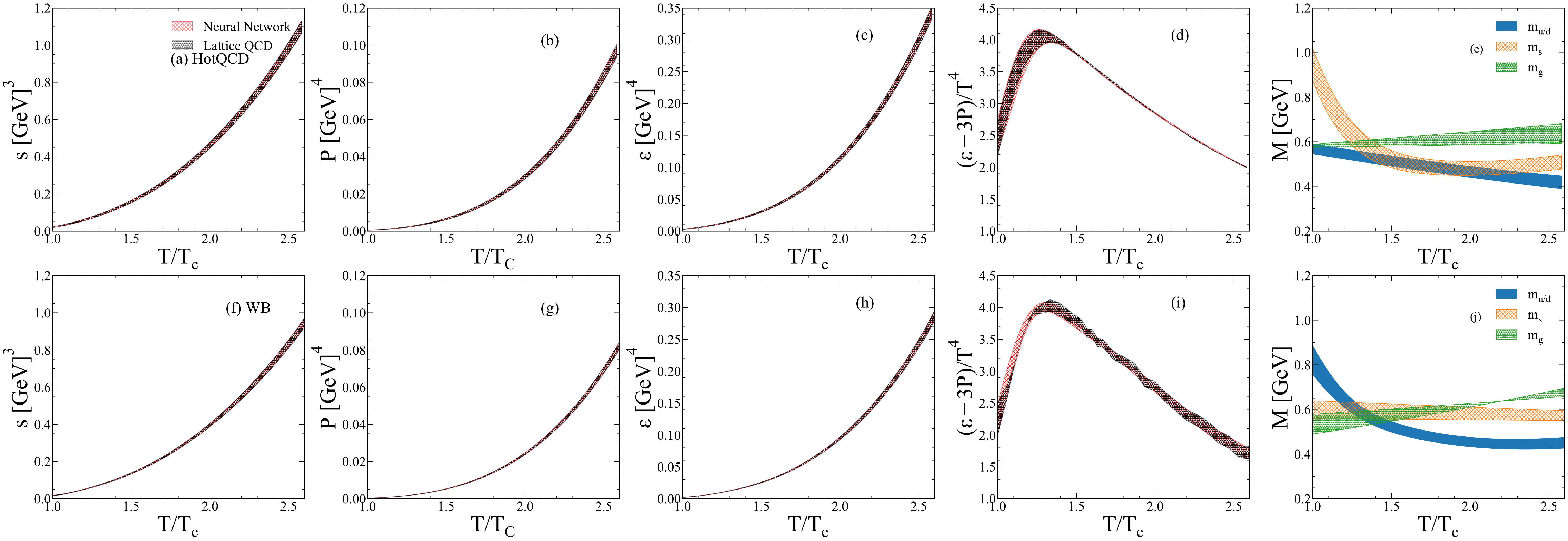}
\caption {\label{fig:disf}(Color online) The uncertainties from training data. The red band is the prediction from DNN and the black band is the input data. Panels (a, b, c, d, e) are the entropy, pressure, energy density,trace anomaly and mass of HotQCD; Panels (f, g, h, i, j) are corresponding WB lattice QCD results.}
\label{fig:witherr}
\end{centering}
\end{figure*}

\section{Uncertainty from training data}
Here we discuss the influence of training data error. Two groups of lattice QCD give not only the values of the equation of state, but also their errors. So we choose the maximum and minimum values of energy density and pressure as two different training samples. Fig.~\ref{fig:witherr} shows that the DNN can perfectly reproduce the error band. Due to the mass constraint at high temperature, the masses obtained from two lattice QCD groups have similar temperature dependence behavior. Since we cannot obtain the correlation between errors, it should be noted that such discussion has certain limitations.

\section{Data}

\setlength{\tabcolsep}{4mm}
  \begin{table*}[htbp]
    \centering
    \caption{The data of Fig.~\ref{fig:MC_HotQCD} for WB lattice QCD}
      \begin{tabular}{ccccc|cccc}
      \multicolumn{5}{c}{DNN}               & \multicolumn{4}{c}{WB Lattice QCD} \\
      T GeV     & s ${\rm GeV}^3$ & P ${\rm GeV}^4$ & $\epsilon$ ${\rm GeV}^4$ & $\Delta$ &  s ${\rm GeV}^3$     & P ${\rm GeV}^4$    & $\epsilon$ ${\rm GeV}^4$     & $\Delta$ \\

      0.15  & 0.015869 & 0.0003 & 0.00208 & 2.331505 & 0.0162 & 0.00032 & 0.002101 & 2.2 \\
      0.155 & 0.019823 & 0.000389 & 0.002684 & 2.627415 & 0.019923 & 0.000411 & 0.002667 & 2.49 \\
      0.16  & 0.024449 & 0.000499 & 0.003412 & 2.921094 & 0.024248 & 0.000523 & 0.003355 & 2.77 \\
      0.165 & 0.029745 & 0.000635 & 0.004273 & 3.197151 & 0.029378 & 0.000657 & 0.004225 & 3.06 \\
      0.17  & 0.035686 & 0.000798 & 0.005269 & 3.44239 & 0.03557 & 0.000819 & 0.005262 & 3.32 \\
      0.175 & 0.042229 & 0.000992 & 0.006398 & 3.647098 & 0.042339 & 0.001013 & 0.006415 & 3.53 \\
      0.18  & 0.049332 & 0.001221 & 0.007659 & 3.806081 & 0.049572 & 0.001239 & 0.007705 & 3.69 \\
      0.185 & 0.056954 & 0.001487 & 0.00905 & 3.918831 & 0.057175 & 0.001499 & 0.009101 & 3.81 \\
      0.19  & 0.065067 & 0.001791 & 0.010571 & 3.98819 & 0.065229 & 0.001811 & 0.010608 & 3.88 \\
      0.195 & 0.073658 & 0.002138 & 0.012225 & 4.019443 & 0.073704 & 0.002154 & 0.012247 & 3.88 \\
      0.2   & 0.082728 & 0.002529 & 0.014017 & 4.018441 & 0.0824 & 0.002528 & 0.014032 & 3.88 \\
      0.205 & 0.092287 & 0.002966 & 0.015953 & 3.994003 & 0.092182 & 0.002967 & 0.015966 & 3.85 \\
      0.21  & 0.102357 & 0.003452 & 0.018043 & 3.951925 & 0.102797 & 0.003442 & 0.018028 & 3.81 \\
      0.215 & 0.11297 & 0.003991 & 0.020298 & 3.896527 & 0.113297 & 0.003974 & 0.020256 & 3.77 \\
      0.22  & 0.124151 & 0.004583 & 0.02273 & 3.833413 & 0.123517 & 0.004568 & 0.022699 & 3.7 \\
      0.225 & 0.135927 & 0.005233 & 0.02535 & 3.765637 & 0.135548 & 0.005203 & 0.025321 & 3.63 \\
      0.23  & 0.148324 & 0.005943 & 0.028171 & 3.695145 & 0.148437 & 0.005933 & 0.028264 & 3.57 \\
      0.235 & 0.161363 & 0.006717 & 0.031203 & 3.623436 & 0.160926 & 0.00671 & 0.031108 & 3.51 \\
      0.24  & 0.175064 & 0.007558 & 0.034457 & 3.551437 & 0.175565 & 0.007531 & 0.034505 & 3.45 \\
      0.245 & 0.189443 & 0.008469 & 0.037945 & 3.479744 & 0.189709 & 0.008467 & 0.037832 & 3.38 \\
      0.25  & 0.204518 & 0.009454 & 0.041676 & 3.40871 & 0.204688 & 0.009453 & 0.041797 & 3.32 \\
      0.26  & 0.23681 & 0.011658 & 0.049913 & 3.269429 & 0.237276 & 0.011653 & 0.04981 & 3.21 \\
      0.27  & 0.272045 & 0.014199 & 0.059253 & 3.133989 & 0.271625 & 0.014243 & 0.059521 & 3.09 \\
      0.28  & 0.310305 & 0.017109 & 0.069776 & 3.001273 & 0.309523 & 0.017087 & 0.070071 & 2.98 \\
      0.29  & 0.351637 & 0.020417 & 0.081558 & 2.871256 & 0.351202 & 0.02044 & 0.081337 & 2.89 \\
      0.3   & 0.396086 & 0.024153 & 0.094673 & 2.742544 & 0.3969 & 0.024138 & 0.09477 & 2.78 \\
      0.31  & 0.443688 & 0.028348 & 0.109195 & 2.615084 & 0.443886 & 0.028352 & 0.108975 & 2.67 \\
      0.32  & 0.494494 & 0.033037 & 0.125201 & 2.488283 & 0.494797 & 0.03303 & 0.124781 & 2.56 \\
      0.33  & 0.548553 & 0.038248 & 0.142774 & 2.363463 & 0.549836 & 0.038305 & 0.142311 & 2.44 \\
      0.34  & 0.605935 & 0.044018 & 0.162 & 2.241044 & 0.605282 & 0.043965 & 0.161697 & 2.33 \\
      0.35  & 0.666723 & 0.050378 & 0.182975 & 2.121797 & 0.664563 & 0.050421 & 0.183076 & 2.23 \\
      0.36  & 0.731016 & 0.057366 & 0.205799 & 2.006391 & 0.732499 & 0.057275 & 0.204913 & 2.13 \\
      0.37  & 0.798908 & 0.065013 & 0.230584 & 1.896634 & 0.800317 & 0.065033 & 0.230522 & 2.05 \\
      0.38  & 0.87053 & 0.073354 & 0.257447 & 1.792855 & 0.872465 & 0.073397 & 0.256472 & 1.96 \\
      0.39  & 0.946029 & 0.082435 & 0.286516 & 1.694925 & 0.943172 & 0.082359 & 0.286867 & 1.87 \\
      0.4   & 1.025543 & 0.09229 & 0.317928 & 1.603838 & 1.024 & 0.09216 & 0.31744 & 1.79 \\
      0.41  & 1.109224 & 0.102958 & 0.351824 & 1.519948 & 1.109628 & 0.102858 & 0.35322 & 1.7 \\
      0.42  & 1.197248 & 0.114487 & 0.388357 & 1.442788 & 1.200226 & 0.11451 & 0.388962 & 1.62 \\
      0.43  & 1.289774 & 0.126917 & 0.427686 & 1.372833 & 1.288013 & 0.126838 & 0.42735 & 1.54 \\
      0.44  & 1.386979 & 0.140294 & 0.469977 & 1.309866 & 1.388499 & 0.140179 & 0.468512 & 1.45 \\
      0.445 & 1.437404 & 0.147358 & 0.492287 & 1.280517 & 1.436374 & 0.147444 & 0.494095 & 1.41 \\
      0.45  & 1.489057 & 0.154672 & 0.515404 & 1.253172 & 1.485338 & 0.154594 & 0.516679 & 1.37 \\
      0.455 & 1.541975 & 0.162249 & 0.539349 & 1.227318 & 1.544821 & 0.162437 & 0.540028 & 1.34 \\
      0.46  & 1.596192 & 0.170101 & 0.564147 & 1.202535 & 1.59631 & 0.170143 & 0.564159 & 1.31 \\
      0.465 & 1.651695 & 0.178219 & 0.58982 & 1.179897 & 1.648932 & 0.17813 & 0.589091 & 1.28 \\
      0.47  & 1.708533 & 0.186619 & 0.616391 & 1.158557 & 1.71308 & 0.186404 & 0.61484 & 1.25 \\
      0.475 & 1.766715 & 0.195305 & 0.643885 & 1.138774 & 1.768336 & 0.195482 & 0.641424 & 1.23 \\
      0.48  & 1.82627 & 0.204285 & 0.672324 & 1.120255 & 1.824768 & 0.204374 & 0.674169 & 1.2 \\
      0.485 & 1.887207 & 0.213564 & 0.701732 & 1.103186 & 1.882388 & 0.213577 & 0.702701 & 1.18 \\
      0.49  & 1.949563 & 0.223156 & 0.73213 & 1.087001 & 1.952973 & 0.223098 & 0.73213 & 1.15 \\
      \end{tabular}%
    \label{tab:f1WB}%
  \end{table*}%

 \setlength{\tabcolsep}{4mm}
  \begin{table*}[htbp]
    \centering
    \caption{The data of Fig.~\ref{fig:MC_HotQCD} for HotQCD}
      \begin{tabular}{ccccc|cccc}
      \multicolumn{5}{c}{DNN}               & \multicolumn{4}{c}{HotQCD} \\
      T GeV     & s ${\rm GeV}^3$ & P ${\rm GeV}^4$ & $\epsilon$ ${\rm GeV}^4$ & $\Delta$ &  s ${\rm GeV}^3$     & P ${\rm GeV}^4$   & $\epsilon$ ${\rm GeV}^4$    & $\Delta$ \\
      0.155 & 0.02009 & 0.000419 & 0.002695 & 2.489833 & 0.020019 & 0.000418 & 0.002684 & 2.43 \\
      0.16  & 0.024622 & 0.000531 & 0.003409 & 2.772527 & 0.024583 & 0.00053 & 0.003404 & 2.76 \\
      0.165 & 0.029879 & 0.000667 & 0.004263 & 3.053969 & 0.029871 & 0.000665 & 0.004263 & 3.07 \\
      0.17  & 0.03584 & 0.000831 & 0.005262 & 3.317213 & 0.035859 & 0.000829 & 0.005266 & 3.34 \\
      0.175 & 0.042464 & 0.001026 & 0.006405 & 3.547592 & 0.042503 & 0.001025 & 0.006413 & 3.56 \\
      0.18  & 0.049703 & 0.001256 & 0.00769 & 3.735877 & 0.049751 & 0.001256 & 0.0077 & 3.74 \\
      0.185 & 0.057508 & 0.001524 & 0.009115 & 3.878481 & 0.057554 & 0.001524 & 0.009124 & 3.88 \\
      0.19  & 0.065846 & 0.001832 & 0.010678 & 3.976321 & 0.06588 & 0.001832 & 0.010685 & 3.97 \\
      0.195 & 0.074696 & 0.002183 & 0.012383 & 4.034138 & 0.074712 & 0.002183 & 0.012386 & 4.03 \\
      0.2   & 0.084059 & 0.00258 & 0.014232 & 4.056794 & 0.08405 & 0.00258 & 0.01423 & 4.05 \\
      0.205 & 0.093939 & 0.003025 & 0.016233 & 4.052982 & 0.093911 & 0.003025 & 0.016227 & 4.05 \\
      0.21  & 0.104357 & 0.00352 & 0.018395 & 4.028294 & 0.104316 & 0.00352 & 0.018386 & 4.03 \\
      0.215 & 0.115341 & 0.004069 & 0.020729 & 3.987808 & 0.115294 & 0.004069 & 0.02072 & 3.99 \\
      0.22  & 0.126917 & 0.004675 & 0.023247 & 3.937011 & 0.126874 & 0.004674 & 0.023238 & 3.94 \\
      0.225 & 0.139115 & 0.00534 & 0.025961 & 3.87959 & 0.139085 & 0.005338 & 0.025956 & 3.88 \\
      0.23  & 0.151965 & 0.006067 & 0.028885 & 3.818307 & 0.151953 & 0.006066 & 0.028883 & 3.82 \\
      0.235 & 0.165494 & 0.00686 & 0.032031 & 3.754679 & 0.165502 & 0.006859 & 0.032034 & 3.76 \\
      0.24  & 0.179728 & 0.007723 & 0.035412 & 3.690079 & 0.179755 & 0.007722 & 0.035419 & 3.69 \\
      0.245 & 0.194687 & 0.008659 & 0.03904 & 3.625934 & 0.19473 & 0.008658 & 0.039051 & 3.63 \\
      0.25  & 0.210393 & 0.009671 & 0.042927 & 3.562278 & 0.210447 & 0.009671 & 0.042941 & 3.57 \\
      0.255 & 0.226864 & 0.010764 & 0.047087 & 3.499359 & 0.226921 & 0.010764 & 0.047101 & 3.5 \\
      0.26  & 0.244116 & 0.011941 & 0.051529 & 3.437222 & 0.244169 & 0.011941 & 0.051543 & 3.44 \\
      0.265 & 0.262162 & 0.013206 & 0.056267 & 3.375994 & 0.262205 & 0.013207 & 0.056278 & 3.38 \\
      0.27  & 0.281017 & 0.014564 & 0.061311 & 3.315612 & 0.281046 & 0.014564 & 0.061318 & 3.32 \\
      0.275 & 0.300697 & 0.016019 & 0.066673 & 3.255257 & 0.300703 & 0.016018 & 0.066675 & 3.26 \\
      0.28  & 0.321204 & 0.017573 & 0.072364 & 3.196187 & 0.321191 & 0.017573 & 0.072361 & 3.2 \\
      0.285 & 0.342552 & 0.019231 & 0.078396 & 3.137961 & 0.342524 & 0.019232 & 0.078388 & 3.14 \\
      0.29  & 0.364762 & 0.021 & 0.084781 & 3.079346 & 0.364715 & 0.021 & 0.084768 & 3.08 \\
      0.295 & 0.387831 & 0.022881 & 0.091529 & 3.022127 & 0.387777 & 0.02288 & 0.091514 & 3.02 \\
      0.3   & 0.411781 & 0.02488 & 0.098654 & 2.964755 & 0.411722 & 0.024879 & 0.098638 & 2.96 \\
      0.305 & 0.436613 & 0.026999 & 0.106168 & 2.908646 & 0.436561 & 0.026999 & 0.106152 & 2.91 \\
      0.31  & 0.462351 & 0.029246 & 0.114082 & 2.852445 & 0.462307 & 0.029246 & 0.114069 & 2.85 \\
      0.315 & 0.489 & 0.031624 & 0.122411 & 2.797008 & 0.488971 & 0.031624 & 0.122402 & 2.79 \\
      0.32  & 0.516577 & 0.034139 & 0.131166 & 2.741809 & 0.516563 & 0.034137 & 0.131163 & 2.74 \\
      0.325 & 0.545086 & 0.036791 & 0.140362 & 2.687891 & 0.545096 & 0.036791 & 0.140365 & 2.69 \\
      0.33  & 0.574551 & 0.03959 & 0.150012 & 2.634342 & 0.57458 & 0.03959 & 0.150022 & 2.63 \\
      0.335 & 0.604979 & 0.042538 & 0.16013 & 2.581809 & 0.605026 & 0.042538 & 0.160145 & 2.58 \\
      0.34  & 0.636389 & 0.045641 & 0.170731 & 2.529872 & 0.636446 & 0.045642 & 0.17075 & 2.53 \\
      0.345 & 0.668793 & 0.048904 & 0.18183 & 2.478887 & 0.668855 & 0.048904 & 0.181851 & 2.48 \\
      0.35  & 0.702206 & 0.052331 & 0.193441 & 2.428886 & 0.702266 & 0.052332 & 0.193461 & 2.43 \\
      0.355 & 0.736641 & 0.055927 & 0.205581 & 2.380097 & 0.736695 & 0.055929 & 0.205598 & 2.38 \\
      0.36  & 0.77213 & 0.059701 & 0.218265 & 2.331572 & 0.772158 & 0.0597 & 0.218276 & 2.33 \\
      0.365 & 0.808664 & 0.063652 & 0.23151 & 2.284855 & 0.808673 & 0.063652 & 0.231513 & 2.29 \\
      0.37  & 0.846273 & 0.067789 & 0.245332 & 2.23916 & 0.846259 & 0.067789 & 0.245327 & 2.24 \\
      0.375 & 0.884972 & 0.072116 & 0.259749 & 2.194708 & 0.884938 & 0.072117 & 0.259735 & 2.2 \\
      0.38  & 0.924779 & 0.076639 & 0.274777 & 2.151375 & 0.924729 & 0.07664 & 0.274757 & 2.15 \\
      0.385 & 0.965719 & 0.081367 & 0.290434 & 2.108831 & 0.965655 & 0.081366 & 0.290412 & 2.11 \\
      0.39  & 1.007793 & 0.0863 & 0.30674 & 2.06792 & 1.007739 & 0.086299 & 0.306719 & 2.07 \\
      0.395 & 1.051025 & 0.091444 & 0.323711 & 2.028392 & 1.051003 & 0.091445 & 0.323701 & 2.03 \\
      0.4   & 1.095446 & 0.096812 & 0.341367 & 1.989541 & 1.095468 & 0.096811 & 0.341376 & 1.99 \\
      \end{tabular}%
    \label{tab:f1HotQCD}%
  \end{table*}%
   \setlength{\tabcolsep}{1mm}
  \begin{table*}[htbp]
    \centering
    \caption{The data of Fig.~\ref{fig:MC_WB}}
      \begin{tabular}{ccccccc|cccccccr}
      \multicolumn{7}{c}{WB lattice QCD (unit: GeV)}                    & \multicolumn{7}{c}{HotQCD (unit: GeV)} \\
      T     & $m_s$ & $m_g$ & $m_{u/d}$ & $\sigma(m_s$) & $\sigma(m_g)$ & $\sigma(m_{u/d})$ & T     & $m_s$& $m_g$ & $m_{u/d}$& $\sigma(m_s)$ & $\sigma(m_g)$ & $\sigma(m_{u/d})$ \\
      0.15  & 0.74359 & 0.63596 & 0.658578 & 0.173052 & 0.068293 & 0.096579 & 0.155 & 0.942646 & 0.590844 & 0.566161 & 0.040658 & 0.025188 & 0.011345 \\
      0.155 & 0.713892 & 0.637146 & 0.643442 & 0.143633 & 0.067284 & 0.084901 & 0.16  & 0.882794 & 0.59166 & 0.562672 & 0.032962 & 0.024656 & 0.011022 \\
      0.16  & 0.68715 & 0.638338 & 0.629033 & 0.117463 & 0.066269 & 0.074024 & 0.165 & 0.82945 & 0.592479 & 0.559202 & 0.027144 & 0.024123 & 0.010705 \\
      0.165 & 0.663217 & 0.639535 & 0.615374 & 0.094473 & 0.065249 & 0.06398 & 0.17  & 0.78213 & 0.5933 & 0.555751 & 0.022776 & 0.02359 & 0.010393 \\
      0.17  & 0.641914 & 0.640738 & 0.602486 & 0.074635 & 0.064222 & 0.054803 & 0.175 & 0.740328 & 0.594124 & 0.552319 & 0.019505 & 0.023056 & 0.010086 \\
      0.175 & 0.623041 & 0.641946 & 0.590381 & 0.058021 & 0.06319 & 0.046529 & 0.18  & 0.703528 & 0.594951 & 0.548905 & 0.017056 & 0.022521 & 0.009784 \\
      0.18  & 0.606387 & 0.643159 & 0.579064 & 0.044917 & 0.062153 & 0.039199 & 0.185 & 0.671234 & 0.595782 & 0.545509 & 0.015224 & 0.021986 & 0.009487 \\
      0.185 & 0.591742 & 0.644378 & 0.568533 & 0.035914 & 0.061109 & 0.032861 & 0.19  & 0.642973 & 0.596615 & 0.542132 & 0.013856 & 0.021451 & 0.009194 \\
      0.19  & 0.5789 & 0.645601 & 0.558773 & 0.031658 & 0.06006 & 0.027568 & 0.195 & 0.6183 & 0.597451 & 0.538774 & 0.012839 & 0.020915 & 0.008906 \\
      0.195 & 0.567663 & 0.646831 & 0.549762 & 0.031893 & 0.059005 & 0.023378 & 0.2   & 0.596807 & 0.59829 & 0.535434 & 0.012086 & 0.020379 & 0.008623 \\
      0.2   & 0.55785 & 0.648065 & 0.541463 & 0.035099 & 0.057944 & 0.02033 & 0.205 & 0.578119 & 0.599132 & 0.532113 & 0.011528 & 0.019842 & 0.008344 \\
      0.205 & 0.549293 & 0.649305 & 0.533832 & 0.039635 & 0.056877 & 0.018402 & 0.21  & 0.561898 & 0.599978 & 0.528809 & 0.01111 & 0.019305 & 0.00807 \\
      0.21  & 0.541844 & 0.650551 & 0.526817 & 0.044497 & 0.055805 & 0.01747 & 0.215 & 0.547838 & 0.600826 & 0.525524 & 0.010791 & 0.018767 & 0.0078 \\
      0.215 & 0.535369 & 0.651802 & 0.520363 & 0.049196 & 0.054726 & 0.017314 & 0.22  & 0.535667 & 0.601678 & 0.522258 & 0.010537 & 0.018229 & 0.007534 \\
      0.22  & 0.529751 & 0.653058 & 0.51441 & 0.053525 & 0.053642 & 0.017676 & 0.225 & 0.525142 & 0.602533 & 0.519009 & 0.010325 & 0.017691 & 0.007272 \\
      0.225 & 0.52489 & 0.65432 & 0.508904 & 0.057405 & 0.052552 & 0.018333 & 0.23  & 0.516051 & 0.603391 & 0.515779 & 0.010137 & 0.017153 & 0.007015 \\
      0.23  & 0.520697 & 0.655587 & 0.503795 & 0.060822 & 0.051456 & 0.019122 & 0.235 & 0.508208 & 0.604252 & 0.512567 & 0.009962 & 0.016614 & 0.006761 \\
      0.235 & 0.517096 & 0.65686 & 0.499038 & 0.063787 & 0.050354 & 0.019941 & 0.24  & 0.501451 & 0.605116 & 0.509373 & 0.00979 & 0.016075 & 0.006511 \\
      0.24  & 0.514023 & 0.658138 & 0.494596 & 0.066326 & 0.049246 & 0.020727 & 0.245 & 0.49564 & 0.605984 & 0.506197 & 0.009618 & 0.015537 & 0.006266 \\
      0.245 & 0.51142 & 0.659422 & 0.490436 & 0.068466 & 0.048132 & 0.021447 & 0.25  & 0.490658 & 0.606855 & 0.503039 & 0.009442 & 0.014998 & 0.006024 \\
      0.25  & 0.509238 & 0.660711 & 0.486532 & 0.070237 & 0.047012 & 0.022082 & 0.255 & 0.486402 & 0.607729 & 0.499899 & 0.009261 & 0.014459 & 0.005786 \\
      0.26  & 0.505976 & 0.663306 & 0.479407 & 0.072782 & 0.044755 & 0.023081 & 0.26  & 0.482788 & 0.608607 & 0.496777 & 0.009074 & 0.013919 & 0.005552 \\
      0.27  & 0.503958 & 0.665922 & 0.473085 & 0.074157 & 0.042474 & 0.023728 & 0.265 & 0.479745 & 0.609488 & 0.493673 & 0.00888 & 0.01338 & 0.005321 \\
      0.28  & 0.502976 & 0.668561 & 0.46748 & 0.074523 & 0.040168 & 0.024063 & 0.27  & 0.477212 & 0.610373 & 0.490587 & 0.008679 & 0.012841 & 0.005094 \\
      0.29  & 0.502872 & 0.671221 & 0.462549 & 0.074005 & 0.037839 & 0.024122 & 0.275 & 0.47514 & 0.611261 & 0.487519 & 0.00847 & 0.012302 & 0.00487 \\
      0.3   & 0.503528 & 0.673904 & 0.458279 & 0.072694 & 0.035486 & 0.023931 & 0.28  & 0.473488 & 0.612152 & 0.484468 & 0.008252 & 0.011763 & 0.00465 \\
      0.31  & 0.504855 & 0.676609 & 0.45468 & 0.070659 & 0.033109 & 0.023503 & 0.285 & 0.472223 & 0.613047 & 0.481435 & 0.008025 & 0.011224 & 0.004433 \\
      0.32  & 0.506785 & 0.679336 & 0.451773 & 0.06795 & 0.030708 & 0.022845 & 0.29  & 0.471315 & 0.613946 & 0.478419 & 0.007787 & 0.010685 & 0.00422 \\
      0.33  & 0.509259 & 0.682086 & 0.449582 & 0.064607 & 0.028283 & 0.02196 & 0.295 & 0.470742 & 0.614848 & 0.475421 & 0.007538 & 0.010147 & 0.00401 \\
      0.34  & 0.512228 & 0.684857 & 0.44813 & 0.060668 & 0.025835 & 0.020851 & 0.3   & 0.470483 & 0.615754 & 0.472441 & 0.007275 & 0.009609 & 0.003803 \\
      0.35  & 0.515649 & 0.687651 & 0.447434 & 0.056167 & 0.023364 & 0.019523 & 0.305 & 0.470522 & 0.616663 & 0.469478 & 0.006999 & 0.009071 & 0.0036 \\
      0.36  & 0.519478 & 0.690467 & 0.447499 & 0.05114 & 0.02087 & 0.017983 & 0.31  & 0.470843 & 0.617575 & 0.466532 & 0.006708 & 0.008534 & 0.003401 \\
      0.37  & 0.523674 & 0.693305 & 0.448322 & 0.045627 & 0.018355 & 0.01624 & 0.315 & 0.471435 & 0.618492 & 0.463604 & 0.006401 & 0.007998 & 0.003204 \\
      0.38  & 0.528199 & 0.696166 & 0.449888 & 0.039669 & 0.01582 & 0.014306 & 0.32  & 0.472286 & 0.619412 & 0.460693 & 0.006079 & 0.007462 & 0.003011 \\
      0.39  & 0.533014 & 0.699049 & 0.452174 & 0.033309 & 0.013269 & 0.012196 & 0.325 & 0.473384 & 0.620336 & 0.457799 & 0.00574 & 0.006927 & 0.002822 \\
      0.4   & 0.538081 & 0.701954 & 0.455143 & 0.026597 & 0.010707 & 0.009926 & 0.33  & 0.474719 & 0.621264 & 0.454922 & 0.005385 & 0.006393 & 0.002637 \\
      0.41  & 0.543361 & 0.704881 & 0.458754 & 0.019589 & 0.008151 & 0.007516 & 0.335 & 0.476282 & 0.622196 & 0.452063 & 0.005015 & 0.005861 & 0.002455 \\
      0.42  & 0.54882 & 0.70783 & 0.462958 & 0.012366 & 0.005641 & 0.004987 & 0.34  & 0.478064 & 0.623131 & 0.44922 & 0.00463 & 0.00533 & 0.002278 \\
      0.43  & 0.55442 & 0.710802 & 0.467703 & 0.005215 & 0.003347 & 0.002371 & 0.345 & 0.480055 & 0.62407 & 0.446395 & 0.004234 & 0.004802 & 0.002106 \\
      0.44  & 0.560128 & 0.713796 & 0.472932 & 0.004015 & 0.002233 & 0.000602 & 0.35  & 0.482245 & 0.625014 & 0.443586 & 0.00383 & 0.004276 & 0.001938 \\
      0.445 & 0.563012 & 0.715301 & 0.475711 & 0.007472 & 0.002722 & 0.001839 & 0.355 & 0.484625 & 0.625961 & 0.440794 & 0.003421 & 0.003755 & 0.001777 \\
      0.45  & 0.565911 & 0.716812 & 0.478588 & 0.01125 & 0.003686 & 0.003214 & 0.36  & 0.487186 & 0.626912 & 0.43802 & 0.003018 & 0.003241 & 0.001623 \\
      0.455 & 0.56882 & 0.718328 & 0.481558 & 0.015123 & 0.004855 & 0.004614 & 0.365 & 0.489917 & 0.627867 & 0.435261 & 0.002633 & 0.002736 & 0.001477 \\
      0.46  & 0.571737 & 0.71985 & 0.484612 & 0.019042 & 0.006117 & 0.006027 & 0.37  & 0.49281 & 0.628826 & 0.43252 & 0.002288 & 0.002247 & 0.001341 \\
      0.465 & 0.574657 & 0.721377 & 0.487744 & 0.022984 & 0.007428 & 0.007447 & 0.375 & 0.495854 & 0.629789 & 0.429795 & 0.002018 & 0.001788 & 0.001219 \\
      0.47  & 0.577578 & 0.72291 & 0.490946 & 0.026939 & 0.008768 & 0.008871 & 0.38  & 0.499039 & 0.630755 & 0.427087 & 0.001872 & 0.001386 & 0.001112 \\
      0.475 & 0.580495 & 0.724449 & 0.494212 & 0.030898 & 0.01013 & 0.010294 & 0.385 & 0.502355 & 0.631726 & 0.424395 & 0.001893 & 0.001107 & 0.001026 \\
      0.48  & 0.583407 & 0.725992 & 0.497534 & 0.034856 & 0.011508 & 0.011714 & 0.39  & 0.505791 & 0.632702 & 0.42172 & 0.002086 & 0.001053 & 0.000965 \\
      0.485 & 0.58631 & 0.727541 & 0.500907 & 0.038807 & 0.012898 & 0.013128 & 0.395 & 0.509339 & 0.633681 & 0.419061 & 0.00242 & 0.001252 & 0.000932 \\
      0.49  & 0.589202 & 0.729096 & 0.504325 & 0.042747 & 0.014298 & 0.014531 & 0.4   & 0.512988 & 0.634665 & 0.416418 & 0.002854 & 0.001614 & 0.000931 \\
      \end{tabular}%
    \label{tab:MCmass}%
  \end{table*}%
\end{appendix}

\end{document}